\documentclass[prl,aps,twocolumn,floatfix,amssymb,showpacs]{revtex4}
\usepackage{epsfig}
\def\bc{\begin{center}}
\def\ec{\end{center}}
\def\be{\begin{equation}}
\def\ee{\end{equation}}
\renewcommand{\vec}[1]{\mbox{\boldmath$#1$}}
\begin{document}
\title{Composite fermion theory of correlated electrons in semiconductor quantum 
dots in high magnetic fields}
\author{Gun Sang Jeon, Chia-Chen Chang, and Jainendra K. Jain}
\affiliation{Department of Physics, 104 Davey Laboratory,
The Pennsylvania State University, University Park, Pennsylvania 16802}
\date{\today}

\begin{abstract}
Interacting electrons in a semiconductor quantum dot 
at strong magnetic fields exhibit a rich set of states, including correlated
quantum fluids and crystallites of various symmetries.
We develop in this paper a perturbative scheme based on the correlated 
basis functions of the composite-fermion theory,
that allows a systematic improvement of the wave functions and the energies
for low-lying eigenstates.  For a test of the method, we study systems
for which exact results are known, and find that practically 
exact answers are obtained for the ground state wave function, ground
state energy, excitation gap, and the pair correlation function.
We show how the perturbative scheme helps resolve the 
subtle physics of competing orders in certain anomalous cases. 
\end{abstract}
\pacs{PACS:73.43.-f,71.10.Pm}
\maketitle

There is a strong motivation for developing theoretical tools  
for obtaining a precise quantitative description  
of interacting electrons in confined geometries, for example in
a semiconductor quantum dot, because of their possible 
relevance to future technology~\cite{qd}.  
Exact diagonalization is possible in some limits but restricted to 
very small numbers of electrons, and does not give insight into the 
underlying physics.  For larger systems, one must 
necessarily appeal to approximate methods.   
The standard Hartree-Fock or density functional 
type methods provide useful insight, but are often not very accurate
for these strongly correlated systems.
The aim of this paper is to demonstrate that a practically exact 
quantitative description is possible for a model quantum dot system,
facilitated by the ability to construct low-energy correlated 
basis functions.

Our concern will be with the solution of 
\be
H=\sum_j\frac{1}{2m_b}\left(\vec{p}_j+\frac{e}{c}\vec{A}_j\right)^2
+\sum_j\frac{m_b}{2} \omega_0^2 r_j^2 +
\sum_{j<k} \frac{e^2}{\epsilon r_{jk}}
\ee
which contains $N$ interacting electrons in two dimensions, confined 
by a parabolic potential and subjected to a magnetic field.
The parameter $m_b$ is the band mass of the electron, $\omega_0$
is a measure of the strength of the confinement, $\epsilon$ is
the dielectric constant of the host semiconductor, and $r_{jk}=
|\vec{r}_j-\vec{r}_k|$.  We will consider  
the limit of a large magnetic field ($\omega_c = eB/m_b c \gg \omega_0$), 
when it is a good approximation to take electrons to 
be confined to the lowest Landau level (LL).
In that limit, the energy eigenvalues have the form  
$E(L)=E_c(L) + V(L)$
where the contribution from the confinement potential 
is explicitly known as a function of the total angular momentum $L$:
$E_{c}(L)=(\hbar/2)[\Omega-\omega_{c}] L$,
with $\Omega^2=\omega_{c}^2+4\omega_{0}^2$,
and $V(L)$ is the interaction energy of electrons
without confinement, but with the magnetic length replaced by
an effective magnetic length given by $\ell\equiv
\sqrt{\hbar/m_b\Omega}$.
Thus, the problem is reduced to finding the interaction energy 
$V$ (which will be quoted below in units of $e^2/\epsilon \ell$) as 
a function of the angular momentum $L$.
Exact results, known for a range of $N$ and $L$ values 
from a numerical diagonalization of the Hamiltonian,
provide a rigorous and unbiased benchmark for any theoretical approach. 
Exact studies have shown~\cite{Seki} a correlated liquid like state at 
small $L$, but a crystallite at relatively large $L$, 
as may be expected from the fact that the system becomes more 
and more classical as $L$ increases.  (The ground state in 
the classical limit is a crystal~\cite{classical}.)
Hartree-Fock studies have been performed for the quantum 
crystallite~\cite{Muller,Landman}.

\begin{figure}
\centerline{\epsfig{file=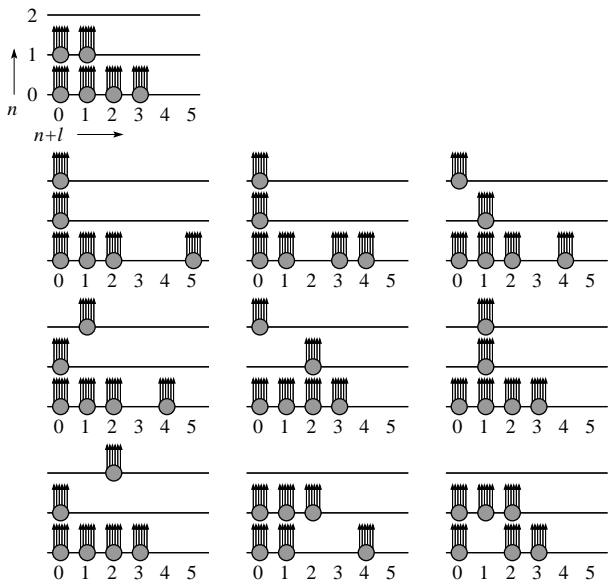,width=8cm}}
\caption{Schematic depiction of Slater determinant basis states for $N=6$ electrons 
at $L=95$, which maps into $L^*=5$ with $2p=6$.  The single 
electron orbitals at $L^*=5$ are labeled by two quantum numbers, the LL index  
$n=0,1,...$,  and the angular momentum 
$l=-n,-n+1,...$  The x-axis labels $n+l$ and the y-axis $n$.  
The dots show the occupied orbitals forming the 
Slater determinants $\Phi^{L^*=5}_{\alpha}$ relevant up to the first 
order.  The state shown at the top left has the lowest kinetic energy 
(if the kinetic energy is measured relative to the lowest Landau level, then, in units of 
the cyclotron energy, the total kinetic energy of this state is two).
The other nine 
states have one higher unit of kinetic energy.  The basis states $\Psi^{L=95}_\alpha$
are obtained according to Eq.~(\ref{eq:CFbasis}), through multiplication by  
$\prod_{j<k}(z_j-z_k)^6$, which converts electrons into composite fermions
carrying six vortices.  That is shown schematically by six arrows on each 
dot.  The single state at the top is relevant at the zeroth order, and all 
ten basis states are employed at the first order.  
(In fact, there are a total of 12 linearly independent states $\Phi_{\alpha}^{L^*}$  
at the first order for $L^*=5$, but they produce only ten linearly 
independent states $\Psi_{\alpha}^L$ at $L=95$.)
\label{fig4}
} 
\end{figure} 

We will apply the composite fermion (CF) theory~\cite{Jain}
to quantum dot states~\cite{Kawamura,JK,Cappelli}.  The central idea is a mapping 
between strongly interacting electrons at angular momentum $L$ and weakly interacting  
electrons at $L^*\equiv L-pN(N-1)$.  In particular, a correlated basis 
$\{\Psi^L_\alpha\}$ for the low energy states of interacting 
electrons at $L$ can be constructed from the trivial, orthonormal Slater 
determinant basis for non-interacting electrons at $L^*$, denoted by 
$\{\Phi^{L^*}_\alpha\}$, in the following manner:
\begin{equation} \label{eq:CFbasis}
\Psi^L_\alpha={\cal P} \prod_{j<k}(z_j-z_k)^{2p} \Phi^{L^*}_\alpha\;. 
\end{equation}
Here, $z_j=x_j-iy_j$ denotes the position of the $j$th electron,
$2p$ is the vorticity of composite fermions,
and ${\cal P}$ indicates projection into the lowest LL.
(Electrons at $L^*$ in general occupy several Landau levels.)
The symbol $\alpha=1, 2, \cdots, D^*$ labels the $D^*$ Slater 
determinants included in the study.  
In general, the basis $\{\Psi^L_\alpha\}$ is not 
linearly independent, so its dimension, $D_{\rm CF}$, may not be equal to  
$D^*$ ($D_{\rm CF}\leq D^*$).  
The advantage of working with the correlated CF basis is that 
$D_{\rm CF}$ is drastically smaller than $D_{\rm ex}$, the dimension
of the lowest LL Fock space at $L$ (which is also the 
dimension of the matrix that must be diagonalized
for obtaining exact results).  Fig.~(\ref{fig4}) illustrates some basis 
functions at $L=95$.

At a back-of-the-envelope level, one can compare the exact interaction
energy at $L$ to the kinetic energy of {\em free} fermions at $L^*$, with the 
cyclotron energy treated as an adjustable parameter~\cite{Kawamura}.  That 
reproduces the qualitative behavior for the $L$ dependence of the exact 
energy for small $L$~\cite{Kawamura}, but discrepancies are known to appear 
at larger $L$~\cite{Seki,Landman}.

For a more substantive test of the theory, it is necessary to obtain 
the energy spectrum by diagonalizing the Hamiltonian of Eq.~1 
in the correlated basis functions of Eq.~(\ref{eq:CFbasis}).  
The CF-quasi-Landau level mixing is treated as a small parameter, and 
completely suppressed at the simplest approximation, which we refer to as 
the zeroth order approximation.
Here, the correlated basis states at $L$ are obtained 
by restricting the basis $\{\Phi_\alpha\}$ to all states with the lowest
kinetic energy at $L^*$ (with $p$ always chosen so as to give the smallest
dimension).

Diagonalization in the correlated CF basis is technically involved, but 
efficient methods for generating the basis functions as well as all of 
the matrix elements required for Gram-Schmidt orthogonalization and 
diagonalization have been developed using Metropolis Monte Carlo sampling.
We refer the reader to earlier literature for full details~\cite{JK,Mandal}.

\begingroup
\squeezetable
\begin{table}
\begin{ruledtabular}
\begin{tabular}{rrrrrrr}  
$L\phantom{a}$& $V_{\rm ex}\phantom{a}$ &$V_{\rm CF}^{(0)}\phantom{a}$ &
$V_{\rm CF}^{(1)}\phantom{a}$ & $D_{\rm ex}$ & $D_{\rm CF}^{(0)}$ &
$D_{\rm CF}^{(1)}$ \\
\hline
  79 & 2.1570 & 2.1610(2)& 2.1573(3)&   26207  &  4 &  55\\
  80 & 2.1304 & 2.1332(1)& 2.1307(1)&   28009  &  2 &  33\\ 
  81 & 2.1286 & 2.1302(1)& 2.1289(2)&   29941  &  1 &  20\\ 
  82 & 2.1226 & 2.1261(4)& 2.1229(5)&   31943  & 10 &  86\\ 
  83 & 2.1090 & 2.1141(7)& 2.1093(2)&   34085  &  5 &  50\\ 
  84 & 2.0893 & 2.0941(1)& 2.0903(2)&   36308  &  2 &  26\\ 
  85 & 2.0651 & 2.0692(1)& 2.0655(1)&   38677  &  1 &  13\\ 
  86 & 2.0651 & 2.0694(5)& 2.0656(2)&   41134  &  5 &  48\\ 
  87 & 2.0543 & 2.0552(2)& 2.0546(2)&   43752  &  2 &  24\\ 
  88 & 2.0462 & 2.0496(1)& 2.0466(2)&   46461  &  9 &  58\\ 
  89 & 2.0279 & 2.0330(3)& 2.0290(3)&   49342  &  3 &  27\\ 
  90 & 2.0054 & 2.0097(1)& 2.0064(1)&   52327  &  1 &   9\\ 
  91 & 2.0054 & 2.0098(3)& 2.0065(2)&   55491  &  3 &  25\\ 
  92 & 1.9989 & 2.0013(1)& 1.9996(2)&   58767  &  8 &  48\\ 
  93 & 1.9852 & 1.9861(1)& 1.9851(3)&   62239  &  2 &  20\\ 
  94 & 1.9715 & 1.9764(2)& 1.9726(2)&   65827  &  4 &  36\\ 
  95 & 1.9506 & 1.9549(2)& 1.9516(2)&   69624  &  1 &  10\\ 
  96 & 1.9506 & 1.9551(2)& 1.9516(4)&   73551  &  2 &  18\\ 
  97 & 1.9447 & 1.9484(1)& 1.9456(5)&   77695  &  5 &  32\\ 
  98 & 1.9347 & 1.9381(3)& 1.9359(4)&   81979  &  9 &  49\\ 
  99 & 1.9189 & 1.9228(1)& 1.9217(4)&   86499  &  1 &  17\\ 
 100 & 1.9001 & 1.9034(2)& 1.9014(3)&   91164  &  2 &  26\\ 
 101 & 1.9001 & 1.9033(2)& 1.9014(1)&   96079  &  4 &  41\\ 
 102 & 1.8947 & 1.8977(2)& 1.8959(3)&  101155  &  7 &  58\\ 
 103 & 1.8855 & 1.8880(2)& 1.8863(2)&  106491  & 12 &  83\\ 
 104 & 1.8712 & 1.8736(2)& 1.8729(1)&  111999  & 18 & 111\\ 
 105 & 1.8533 & 1.8617(2)& 1.8542(3)&  117788  &  1 &  28\\ 
 106 & 1.8533 & 1.8618(1)& 1.8538(4)&  123755  &  1 &  39\\ 
 107 & 1.8483 & 1.8555(2)& 1.8488(4)&  130019  &  2 &  55\\ 
 108 & 1.8396 & 1.8463(2)& 1.8402(4)&  136479  &  3 &  74\\ 
\end{tabular}                                                      
\end{ruledtabular}                                                 
\caption{
\label{Etable}
Exact ground state energy ($V_{\rm ex}$) and the ground state energy 
obtained from the zeroth ($V_{\rm CF}^{(0)}$) 
and the first-order ($V_{\rm CF}^{(1)}$) CF theory for $N=6$. 
The dimensions of the bases diagonalized are $D_{\rm ex}$, $D_{\rm CF}^{(0)}$
and $D_{\rm CF}^{(1)}$, respectively. 
The statistical uncertainty arising from Monte Carlo sampling is 
given in parentheses. 
}
\end{table}
\endgroup

\renewcommand{\sp}{@{\hspace*{2ex}}}
\begin{table}
\begin{ruledtabular}
\begin{tabular}{\sp r\sp cc\sp|r\sp cc\sp}  
\multicolumn{1}{\sp c}{$L$} & ${\cal O}^{(0)}$ & ${\cal O}^{(1)}$ 
& \multicolumn{1}{ c\sp}{$L$} & ${\cal O}^{(0)}$ & ${\cal O}^{(1)}$ \\
	\hline
95 & 0.902 & 0.988 &102& 0.927 & 0.985\\
96 & 0.892 & 0.988 &103& 0.943 & 0.978\\
97 & 0.898 & 0.989 &104& 0.946 & 0.972\\
98 & 0.908 & 0.985 &105& 0.714 & 0.989\\
99 & 0.767 & 0.859 &106& 0.710 & 0.987\\
100& 0.936 & 0.982 &107& 0.735 & 0.988\\
101& 0.936 & 0.981 &108& 0.781 & 0.990
\end{tabular}                                                      
\end{ruledtabular}                                                 
\caption{
\label{Ctable}
Overlaps between exact ground states and CF ground states obtained
at the zeroth (${\cal O}^{(0)}$) and the first order(${\cal O}^{(1)}$).
The statistical uncertainty from Monte Carlo sampling does not affect 
the first three significant figures. 
}
\end{table}

A diagonalization of the Hamiltonian in the zeroth level basis produces  
energies and wave functions for $D_{\rm CF}^{(0)}$ low-lying states.
The interaction energy and the wave function for the ground state 
will be denoted $V^{(0)}_{CF}$ and $\Psi_{CF}^{(0)}$, respectively.  
We have carried out~\cite{Jeon1} extensive calculations 
for a large range of $L$ for up to ten particles, and found that 
the CF theory reproduces the qualitative behavior of the energy as a function 
of $L$ all the way to 
the largest $L$ for which exact results are known.  We show in Table~\ref{Etable} 
results for $N=6$ electrons in the angular momentum range 
$79\leq L \leq 108$, which spans both liquid and crystal-like ground 
states.  Ref.~\cite{Jeon1} shows a comparison of the exact pair correlation function 
with that calculated 
from  $\Psi_{CF}^{(0)}$ for $L=95$ (where there is a unique CF wave function);
surprisingly, the CF theory, originally intended for the liquid state,
automatically produces also a crystallite at large $L$, even though  
no crystal structure has been put into the theory by hand~\cite{Fertig}.
Table~\ref{Ctable} gives the overlaps defined as:
${\cal O}^{(0)} \equiv {\big|\langle \Psi_{\rm CF}^{(0)} |\Psi_{\rm ex} \rangle \big|}/
{\sqrt{\langle \Psi_{\rm CF}^{(0)} |\Psi_{\rm CF}^{(0)} \rangle
\langle \Psi_{\rm ex} |\Psi_{\rm ex} \rangle }}$.

\begin{figure}
\parbox{1.5in}{
\centerline{\epsfig{file=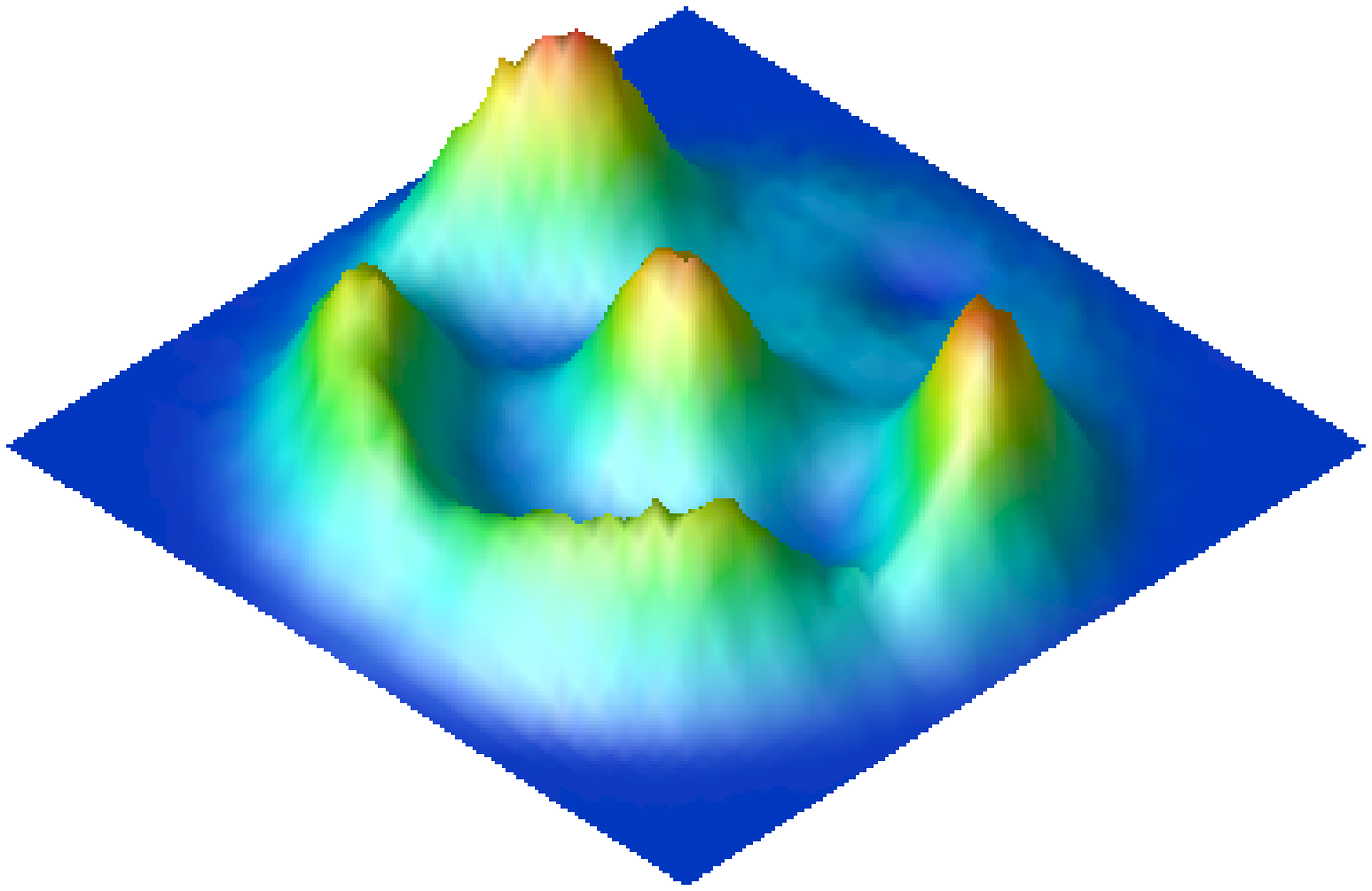,width=1.5in}}
\centerline{(a)}
\epsfig{file=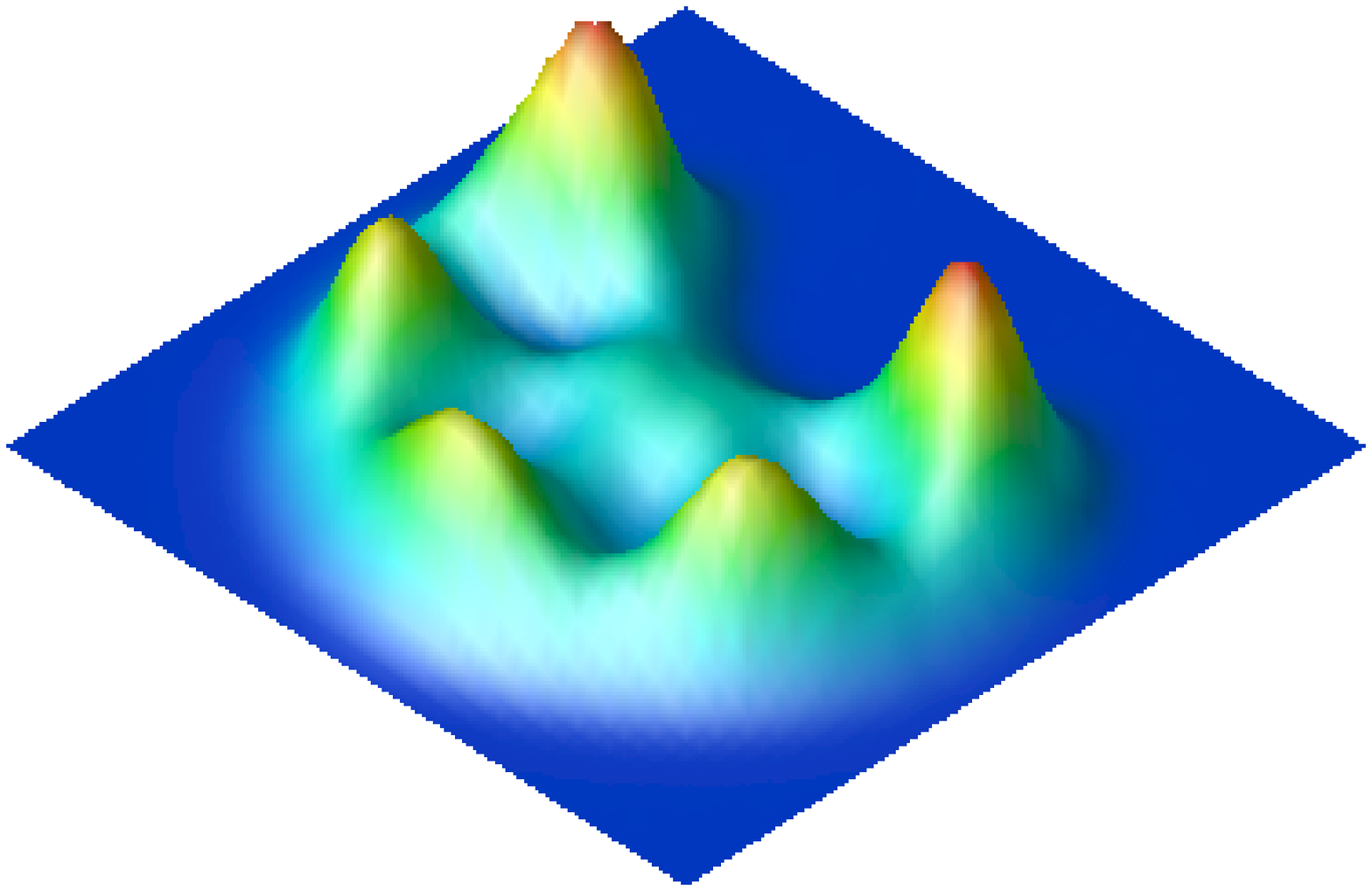,width=1.5in}
\centerline{(c)}
}
\parbox{1.5in}{
\epsfig{file=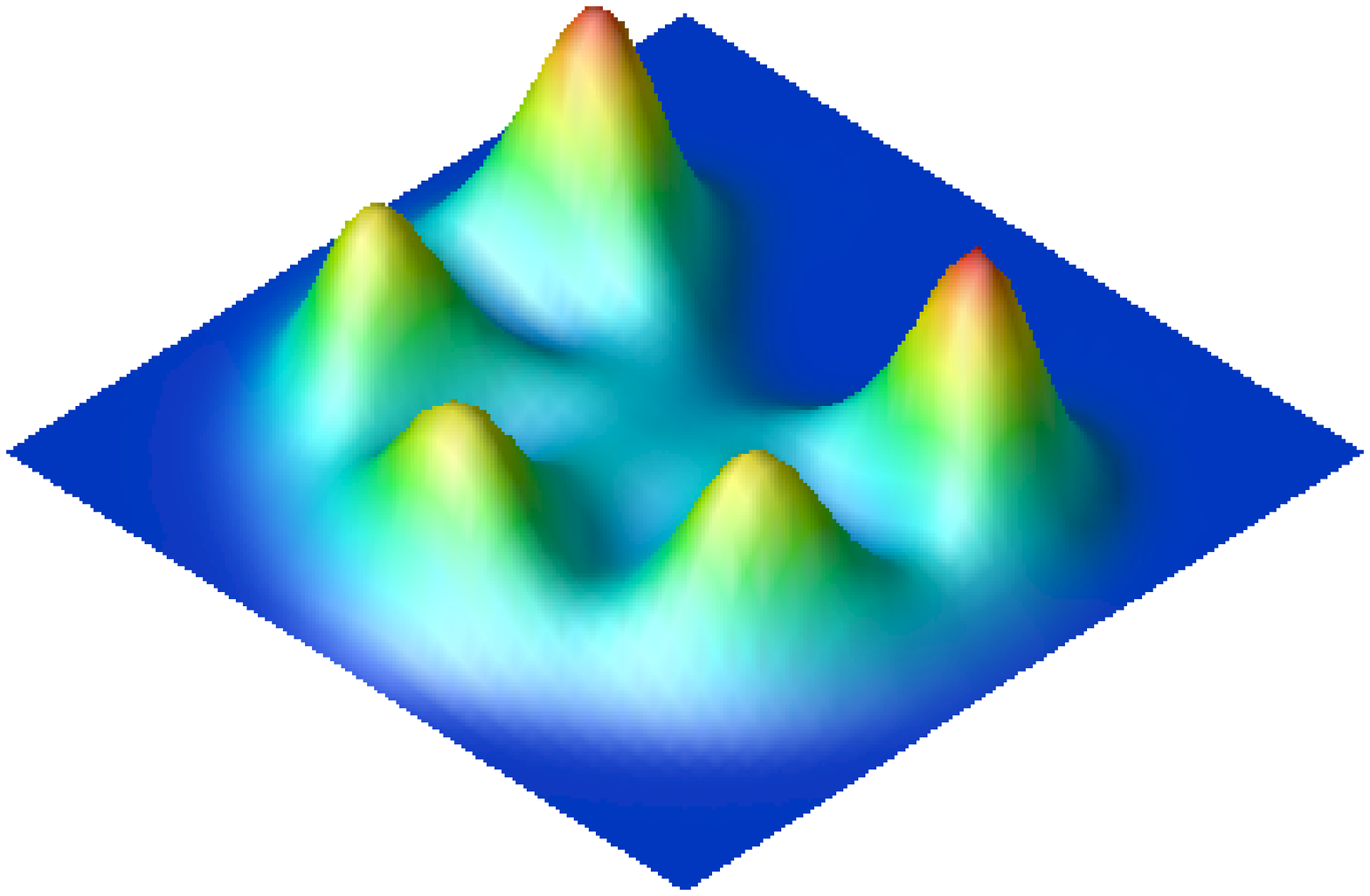,width=1.5in}
\centerline{(b)}
\epsfig{file=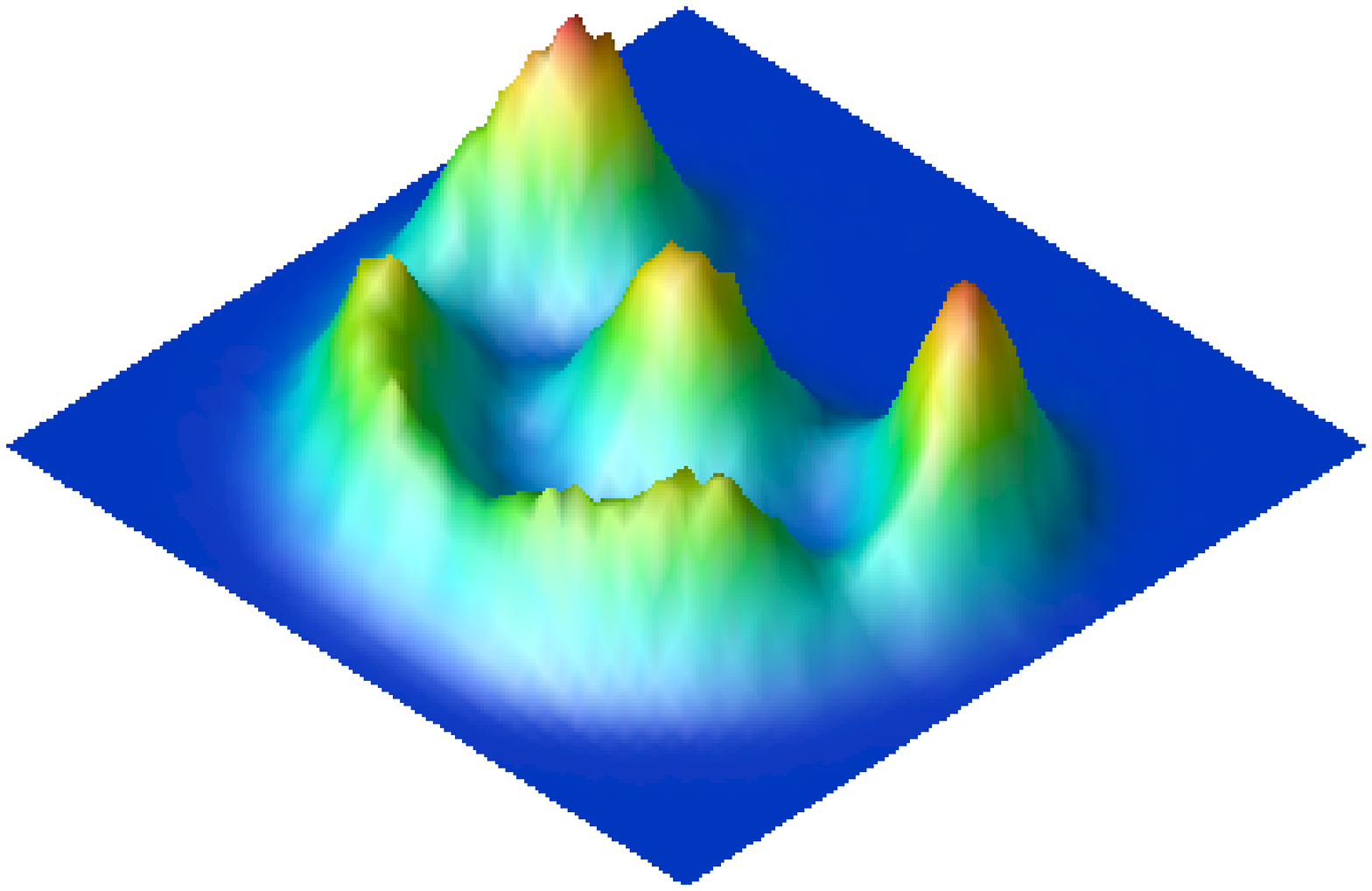,width=1.5in}
\centerline{(d)}
}
\caption{
\label{fig2}
Pair correlation function for $N=6$
electrons at $L=99$.  The position of one particle is fixed on the  
outer ring, coincident with the position of the missing peak.  
The ground state wave function used in the calculation 
is obtained from (a) exact diagonalization; (b) the 
zeroth-order CF theory; (c) the first-order CF theory; (d) the 
second-order CF theory.  The ``noise" in (a) and (d) results from 
the relatively large statistical uncertainty in Monte Carlo because of 
the more complicated wave function.}
\end{figure}

While the zeroth level description is quite good, the following 
deviations from the exact solution may be noted.  (i)  The overlaps are  
in the range 0.70-0.94, which are not as high as the overlaps
($\sim$0.99) for incompressible ground states in the spherical geometry.
(ii) The energies are within 0.5\% of the exact ones, which 
is quite good but could be further improved.
(iii) In the crystallite, the particles are somewhat less strongly
localized in the CF wave function than in the exact ground state.
See Ref.~\cite{Jeon1}. 
(iv)  The symmetry of the crystallite is predicted correctly with the 
exception of $L=99$, where the CF theory predicts
a $(0,6)$ crystallite [Fig.~\ref{fig2}(b)],
that is, with all six particles on
an outer ring, whereas the exact solution shows a $(1,5)$ crystallite
[Fig.~\ref{fig2}(a)],
which has five particles on the outer ring and one at the center.
(v)  A successful theory must explain not only the ground state but 
also excited states, especially the low-energy ones.  We have considered the  
gap between the two lowest eigenstates.
The zeroth-order theory does not give, overall, a satisfactory account of it.
In some instances (e.g., $L=81,85,90,95,99,105,106$ for $N=6$), the 
CF theory gives no information on the gap, because the basis contains only 
a single state here ($D^{(0)}_{CF}=1$); in many other cases, 
the gap predicted by the CF theory is off by a factor of two to three.

These discrepancies have motivated us to 
incorporate CF-quasi-LL mixing perturbatively.  (We stress that CF-quasi-LL mixing 
implies LL mixing at $L^*$, but the basis states at $L$ are, by construction,
always within the lowest LL.) 
At the first order, we include basis states $\Phi^{L^*}$ at $L^*$ with one more unit 
of the kinetic energy, which produces a larger basis at $L$ 
through Eq.~(\ref{eq:CFbasis}).
The basic idea is illustrated for the case of $N=6$ and $L=95$ ($L^*=5$,
$2p=6$) in Fig.~\ref{fig4}.


\begin{figure}[b]
\centerline{\epsfig{file=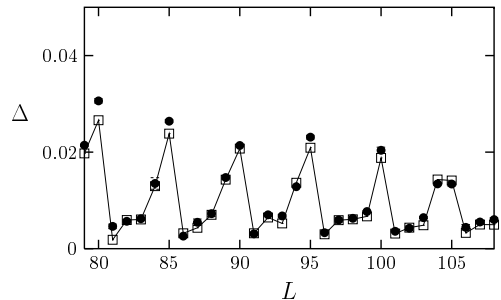,width=6.5cm}}
\caption{
	\label{fig3}
	Comparison of the exact excitation gaps ($\square$) for $N=6$ with
the gaps obtained in the first-order CF theory ({\Large $\bullet$}).}
\end{figure}

In a similar way we have constructed correlated basis functions at each angular
momentum in the range $79\leq L \leq 108$.  As shown in Table~\ref{Etable}
$D_{\rm CF}^{(1)}$, the dimension of the basis  
in the first-order theory is larger than $D_{\rm CF}^{(0)}$
but still far smaller than $D_{\rm ex}$.  A diagonalization of 
the Hamiltonian in this basis produces the ground state energy $V^{(1)}_{CF}$ and 
ground state wave function $\Psi^{(1)}_{CF}$.  Leaving aside $L=99$, which we 
shall discuss separately, the following observations can be made: 
(i) The energies are essentially exact.  As shown in Table~\ref{Etable}, the   
deviation from the exact energy is reduced to $<$0.1\%, in fact, to $<$0.05\% 
in most cases.  (ii) The overlaps from the first-order theory,  
${\cal O}^{(1)} \equiv {\big|\langle \Psi_{\rm CF}^{(1)} |\Psi_{\rm ex} \rangle \big|}/
{\sqrt{\langle \Psi_{\rm CF}^{(1)} |\Psi_{\rm CF}^{(1)} \rangle
\langle \Psi_{\rm ex} |\Psi_{\rm ex} \rangle }}$,
are given in Table~\ref{Ctable}.   They are  
uniformly excellent (0.98-0.99) in the entire $L$ range studied.
(iii) The improvement by the first order perturbation theory is also manifest in the
pair-correlation functions, which are now indistinguishable from the exact ones
at arbitrary $L$.  That is not surprising, given the high overlaps.
(iv) As seen in Fig.~\ref{fig3} the first-order theory reproduces the 
qualitative behavior of the excitation gap as a function of $L$, and also 
gives very good quantitative values.  The maximum gaps are correlated with the
states where a downward cusp appears in the plot of $V(L)$, 
consistent with the higher stability of these ground states.

\begin{table}
\begin{tabular}{cccccc}\hline
$L$ & $D_{ex}$ & $D^{(2)}_{CF}$  &  $V_{ex}$ &  $V^{(2)}_{CF}$  & ${\cal O}^{(2)}$ \\ \hline
99  & 86499    &   76            & 1.9189    &  1.9193(3)       & 0.995 \\ \hline
\end{tabular}
\caption{
        \label{L99table}
Comparison of the second order CF theory with exact results
for the $L=99$ ground state.  $D^{(2)}$ is the dimension of 
the correlated CF basis, $V^{(2)}_{CF}$ is the CF prediction for the 
ground state energy, and ${\cal O}^{(2)}$ is the overlap between the CF and the
exact wave functions.
}
\end{table}

Finally, we discuss the case of $L=99$.
Here, the zeroth order CF theory predicts a wrong symmetry for the 
crystallite [Fig.~\ref{fig2}(b)].  
As seen in Fig.~\ref{fig2}(c), the first order correction also fails to 
recover the correct symmetry.  That is also reflected in the fact 
that the modified ground state of the 
first-order theory yields a relatively small overlap of $\sim 0.86$,
and the energy is off by a relatively large 0.15\%.
A closer inspection of the correlations in Fig.~\ref{fig2}(c) 
reveals a slight broadening of the hexagonal
structure in the outer-ring, combined with an appearance of a small mound at
the center, suggesting that the structure here may be a superposition
of (0,6) and (1,5) crystallites.  
This has motivated us to incorporate the next (second) order corrections.
The basis dimension $D^{(2)}_{CF}$ is now further enlarged,
but Monte Carlo still produces reliable results.
The CF ground state wave function obtained at this level is extremely 
accurate: the pair correlation function shown in  
Fig.~\ref{fig2}(d) compares well to the exact one, and, 
as seen in Table~\ref{L99table}, the energy and 
the overlaps are close to perfect.  The origin of the difficulty can be 
understood from the fact that the (0,6) and the (1,5) crystallites are nearly 
degenerate in the classical limit~\cite{classical}, making them both competitive.

We have focused in this work on cases where the exact results are known,
because the aim was to test the applicability of the CF theory to quantum dots.
The theory can be extended to much larger systems, where exact diagonalization is 
not possible; in such cases, one would need to 
increase the accuracy perturbatively until sufficient convergence is achieved. 
The method developed here should also prove useful for multiple coupled quantum 
dots~\cite{double} and rapidly rotating atomic Bose-Einstein 
condensates~\cite{BEC}.

Partial support by the National Science Foundation under grant
no. DMR-0240458 is gratefully acknowledged.

\end{document}